\documentclass{ptapap}
\usepackage{color}

\usepackage{soul}

\usepackage{amsmath}
\usepackage{amssymb}

\author{Pawe{\l} Zieli{\'n}ski}[OAUW]
\author{{\L }ukasz Wyrzykowski}[OAUW]
\author{Przemys{\l}aw Miko{\l}ajczyk}[IAUWr]
\author{Krzysztof Rybicki}[OAUW]
\author{Zbigniew Ko{\l}aczkowski}[IAUWr,CAMK,*]
\affil[OAUW]{Astronomical Observatory, University of Warsaw, Al. Ujazdowskie 4, 00--478 Warszawa, Poland}
\affil[IAUWr]{Astronomical Institute, University of Wroc{\l}aw, Kopernika 11, 51--622 Wroc{\l}aw, Poland}
\affil[CAMK]{Nicolaus Copernicus Astronomical Center, Polish Academy of Sciences, Bartycka 18, 00--716 Warsaw, Poland}
\affil[*]{deceased}

\title{Towards an automatic processing of CCD images with CPCS 2.0}

\begin{document}

\maketitle

\begin{abstract}
We present a new automatic tool for time-domain astronomy -- the Cambridge Photometric Calibration Server 2.0 -- developed under OPTICON H2020 programme. It has been designed to respond to the need of automated rapid photometric data calibration and dissemination for transient events, primarily from {\it Gaia} space mission. CPCS has been in operation since 2013 and has been used to calibrate around 130\,000 observations of hundreds of transients. We present the status of this tool's development and demonstrate improvements made in the second version. The tests present the ability to combine CCD imaging data from multiple telescopes and a whole variety of instruments. New tool provides science-ready photometric data within minutes from observations in the automatic manner.
\end{abstract}

\section{Motivation}

Within the OPTICON Time-Domain Astronomy Work Package\footnote{http://astro-opticon.org} we have been coordinating the operation of approximately 100 small- and medium-sized telescopes scattered around Europe and beyond. In addition to existing modern, and more often robotic telescopes, OPTICON network of volunteering observers (professionals and amateurs) can provide very useful scientific data and serve as great training facilities for young generations of astronomers. Such network of telescopes and instruments is also of great importance in the era of large sky surveys, like {\it Gaia}
.  It helps to launch follow-up ground-based observations of discovered transients immediately as well as to carry out further data processing.

In order to conduct these observations in an efficient way, one needs to have an automated software dedicated for processing of time-domain data. Therefore, one of the main product of OPTICON TDA network is the Cambridge Photometric Calibration Server (CPCS) which allows fast and automatic calibration of photometric data delivered by the network instruments. Recently, we have been developing the new version of this software -- CPCS 2.0 -- which improves old version in several issues \citep{Zielinski2019}.

\section{Main specification}

\begin{figure}
\vspace{-1.5cm}
\begin{minipage}{0.4\textwidth}
\includegraphics[width=\textwidth]{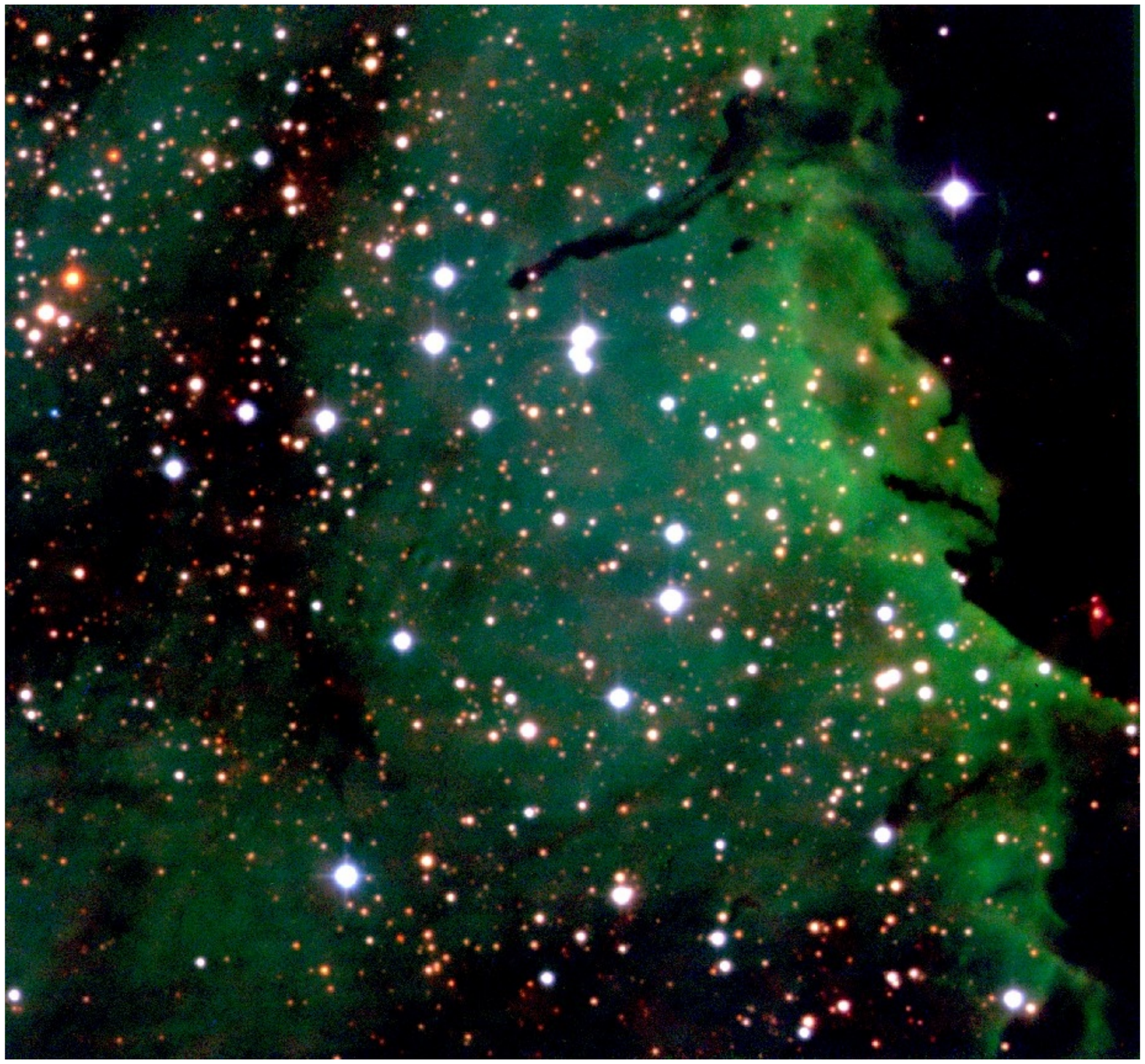}
\end{minipage}
\hspace{1.2cm}
\begin{minipage}{0.39\textwidth}
\includegraphics[width=\textwidth]{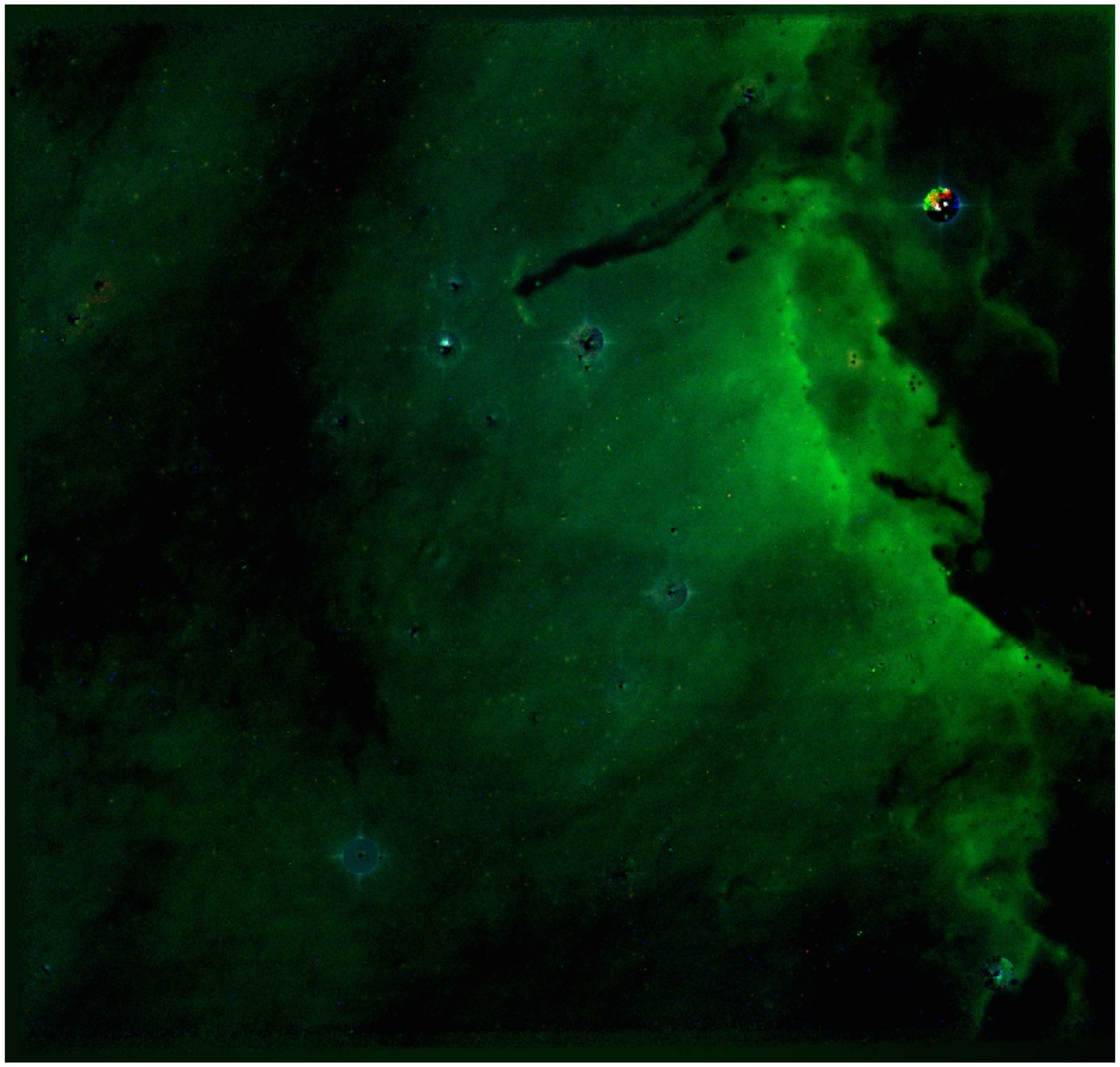}
\end{minipage}
\vspace{-1.5cm}
\caption{{\it Left:} Example image of V1490~Cyg field taken with 60-cm Cassegrain Telescope in Bia{\l}k{\'o}w Observatory, Poland. The image is stacked in V (5$\times$100s), R (5$\times$100s), I (5$\times$90s) filters. FoV $=11\times11$ arcmin. North is up and East is to the left. {\it Right:} The same field, but after CCDPhot astrometric and photometric processing with all stars subtracted.}
\label{fig1}
\end{figure}

Currently, the observers can use the first version of CPCS by uploading simple ASCII files onto the following website: \texttt{http://gsaweb.ast.cam.ac.uk/followup}. It allows to combine data gathered by different instruments in order to provide a photometric light curves. The new version of CPCS is based on CCDPhot package (Mikołajczyk et al., {\it in prep.}) as an engine of automated data processing tool. The example images before and after CCDPhot processing are presented in Fig.~\ref{fig1}. The CPCS 2.0 main features and workflow are as follows:
\begin{itemize}
\item standardization of the headers of FITS
files. The files must be already after bias, dark subtraction, flat-field normalization, etc.;
\item creation of the astrometric catalogues of reference stars in specified sky region by using URAT-1 \citep[][the U.S. Naval Observatory Robotic Astrometric Telescope]{Zacharias2015}, UCAC-4 \citep[][the U.S. Naval Observatory CCD  Astrograph  Catalog]{Zacharias2013}, or USNOB1 \citep[][the U.S. Naval Observatory]{Monet2003} and {\it Gaia}-DR2 \citep{Gaia2018};
\item initial astrometric matching with {\it Gaia}-DR2 catalogue and others;
\item astrophotometry of all stars identified on CCD image with SExtractor \citep[][Source Extractor]{Bertin1996} and SCAMP \citep[][Software for Calibrating AstroMetry and Photometry]{Bertin2006};
\item automatic selection of stars to construct the Point Spread Function (PSF) model;
\item PSF photometry with DAOphot \citep[][the Dominion Astrophysical Observatory package for stellar photometry]{Stetson1987,Stetson1992} for all unsaturated and unblended stars detected in the image (with average precision well below $\sim0.01$ mag);
\item aperture photometry with NEDA routine of DAOphot for these stars;
\item final precision astrometry with {\it Gaia}-DR2 (with precison down to $\sim0.002$'');
\item delivering ASCII file with photometry and astrometry of all stars in the image, as well as FITS file with WCS coordinates (ICRS)\footnote{World Coordinate System, standard of astronomical coordinate system in FITS images};
\item transformation of instrumental magnitudes of selected target to the standard ones by zero-point calibration and colour term application based on ASAS \citep[][the All-Sky Automated Survey]{Pojmanski1997}, SDSS \citep[][the Sloan Digital Sky Survey]{Gunn1998}, PS1 \citep[][the Panoramic Survey Telescope and Rapid Response System]{Chambers2016}, DES \citep[][the Dark Energy Survey]{Abbott2018} and 2MASS \citep[][the Two Micron All Sky Survey]{Skrutskie2006} catalogues;
\item automatic uploading of the results for selected target to the online server in order to produce science-ready light curves.
\end{itemize}

\section{Summary}

This online tool is providing the service to the astronomical community of observers, who are willing to contribute with their observations of objects for which the time-domain aspect is important, e.g., variable stars or various transient phenomena (supernovae, cataclysmic variables, microlensing, etc.). The service is currently used as a central point of Gaia Photometric Science Alerts\footnote{http://gsaweb.ast.cam.ac.uk/alerts/home}, where the data are being stored and calibrated in a homogenous way, so that they can be used for future scientific research (see \citealt{Wyrzykowski2018}). The work under CPCS 2.0 is in progress now, but it should be available via the dedicated website soon. We encourage all interested astronomers to use CPCS for their own research. In order to start using our tool, please contact one of us: {\L}.Wyrzykowski (\texttt{lw@astrouw.edu.pl}) or P. Zieli{\'n}ski (\texttt{pzielinski@astrouw.edu.pl}).

\acknowledgements{We acknowledge the support of a number of observers involved in collecting follow-up photometric data and Gaia Science Alerts Team, Cambridge, UK. We acknowledge the European Union's Horizon 2020 grant No. 730890 (OPTICON) and DAINA NCN grant 2017/27/L/ST9/03221. PM acknowledges support from the NCN grant no. 2016/21/B/ST9/01126. }

\bibliographystyle{ptapap}
\bibliography{zielinski}

\end{document}